# Tipping the Balance: Impact of Class Imbalance Correction on the Performance of Clinical Risk Prediction Models

Running title: Impact of Class Imbalance Correction

Authors: Amalie Koch Andersen[1,2] (MSc); Hadi Mehdizavareh[3] (MSc); Arijit Khan[3, 4] (PhD); Tobias Becher[5,6] (MD); Simone Britsch[5,6] (MD); Markward Britsch[5-10] (PhD); Morten Bøttcher[11,12] (MD, PhD); Simon Winther[11,12] (MD, PhD); Palle Duun Rohde[13] (PhD); Morten Hasselstrøm Jensen[1,2] (PhD); Simon Lebech Cichosz[1*] (PhD)

Author Affiliations:

[1] Medical Informatics and Image Analysis, Department of Health Science and Technology, Aalborg University, Denmark.

[2] Medical & Translational Science, Novo Nordisk A/S, Søborg, Denmark

[3] Department of Computer Science, Aalborg University, Aalborg, Denmark

[4] Department of Computer Science, Bowling Green State University, Bowling Green, Ohio, USA.

[5] Cardiology, Angiology, Haemostaseology, and Medical Intensive Care, Medical Centre Mannheim, Medical Faculty Mannheim, Heidelberg University, Heidelberg, Germany.

[6] European Centre for AngioScience (ECAS), German Centre for Cardiovascular Research (DZHK) partner site Heidelberg/Mannheim, and Centre for Cardiovascular Acute Medicine Mannheim (ZKAM), Medical Centre Mannheim, Medical Faculty Mannheim, Heidelberg University, Heidelberg, Germany.

[7] HMS Analytical Software GmbH, Heidelberg, Germany.

[8] Department of Radiation Oncology, Medical Center Mannheim, Medical Faculty Mannheim, Heidelberg University, Heidelberg, Germany.

[9] DKFZ Hector Cancer Institute, Medical Centre Mannheim, Medical Faculty Mannheim, Heidelberg University, Heidelberg, Germany.

[10] Mannheim Institute for Intelligent Systems in Medicine (MIISM), Heidelberg University, Heidelberg, Germany.

[11] Department of Cardiology, Gødstrup Hospital, Hospitalsparken 15, 7400 Herning, Denmark

[12] Department of Clinical Medicine, Aarhus University, Palle Juul-Jensens Boulevard 99, 8200 Aarhus N

[13] Genomic Medicine, Department of Health Science and Technology, Aalborg University, Denmark

*Corresponding author: Simon Cichosz, simcich@hst.aau.dk, Postal address: Selma Lagerløfs Vej 249, 12-02-048, 9260 Gistrup, Danmark, Phone: (+45) 9940 2020; Fax: (+45) 9815 4008. ORCID: 0000-0002-3484-7571

Words: 3984

# Abstract

**Objective:** Machine-learning–based clinical risk prediction models are increasingly used to support decision-making in healthcare. While class-imbalance correction techniques are commonly applied to improve model performance in settings with rare outcomes, their impact on probabilistic calibration remains insufficiently understood. This study evaluated the effect of widely used resampling strategies on both discrimination and calibration across real-world clinical prediction tasks. **Materials and Methods:** Ten clinical datasets spanning diverse medical domains and including over 600,000 patients (n=605,842) were analyzed. Multiple machine-learning model families, including linear models and several non-linear approaches, were evaluated. Models were trained on the original data and under three commonly used 1:1 class-imbalance correction strategies (SMOTE, RUS, ROS). Performance was assessed on held-out data using discrimination and calibration metrics. **Results:** Across all datasets and model families, resampling had no positive impact on predictive performance. Changes in the Receiver Operating Characteristic Area Under Curve (ROC-AUC) relative to models trained on the original data were small and inconsistent (ROS: -0.002, $p<0.05$; RUS: -0.004, $p>0.05$; SMOTE: -0.01, $p<0.05$), with no resampling strategy demonstrating a systematic improvement. In contrast, resampling in general degraded the calibration performance. Models trained using imbalance correction exhibited higher Brier scores (0.029 to 0.080, $p<0.05$), reflecting poorer probabilistic accuracy, and marked deviations in calibration intercept and slope, indicating systematic distortions of predicted risk despite preserved rank-based performance. **Conclusion:** In a diverse set of real-world clinical prediction tasks, commonly used class-imbalance correction techniques did not provide generalizable improvements in discrimination and were associated with degraded calibration.

# Introduction

In healthcare, machine learning based predictive risk models are increasingly being utilized to guide clinical decisions and to communicate risk of disease or complications [1–4]. Beyond discrimination ability of the models, calibration - the agreement between predicted and actual risk - is critical for patient care [5,6].

Poor calibration can mislead both clinicians and patients, since systematic over- or underestimation of risk may cause unnecessary alarm or anxiety, leading to unwarranted treatment, or give false hope of a successful outcome[7]. In practice, many clinical endpoints are relatively rare, resulting in class imbalance between cases and non-cases[8]. In such settings, conventional metrics like c-statistics (Area Under the Receiver Operating Characteristic Curve (ROC-AUC)) can be misleading. For example, an algorithm may have a high area under the curve yet very low positive predictive value when events are rare. To address this imbalance-correction techniques are often applied in the scientific literature [9–13], but their impact on model performance, especially calibration, is not well understood [14]. Common class-imbalance correction methods include oversampling the minority class (e.g. duplicating cases or using synthetic methods like Synthetic Minority Oversampling Technique (SMOTE)), undersampling the majority class, class weighting in the objective function and cost-sensitive weighting or threshold adjustment [15–20]. These techniques rebalance the training data with the aim of improving metrics like sensitivity or overall accuracy. Such methods have indeed received growing interest [21], because in imbalanced datasets even trivial prediction (always predicting the majority class) can yield deceptively high accuracy. However, shifting the class balance may distort the predicted probabilities and thus harm calibration [21]. Notably, any gain in sensitivity from resampling can often be achieved more simply by moving the decision threshold on a well-calibrated model, rather than by re-training the model on resampled data [21].

Recent studies highlight these trade-offs. Van den Goorbergh *et al.* [21] used simulations of logistic regression and showed that methods like random oversampling, undersampling, or SMOTE produced models with poor calibration - such that the minority-class probabilities were strongly overestimated - and no improvement in discrimination [21]. In their simulations, any apparent balance in sensitivity could be achieved by threshold tuning, and the resampled models remained miscalibrated. They concluded that outcome imbalance is not inherently problematic, and that applying imbalance correction techniques may, in some cases, degrade model performance. Similarly, Carriero *et al.* [14] conducted extensive Monte Carlo simulations across multiple machine learning algorithms. They found that models trained without class correction consistently had equal or better calibration than those with oversampling or undersampling [14]. The calibrated models introduced by balancing methods systematically overestimated risk, and this miscalibration was not fully correctable even with standard recalibration techniques. These findings confirm a broader point, that while imbalance correction can improve certain classification metrics - e.g. by raising sensitivity at a chosen threshold - it often does not increase ROC-AUC and can substantially degrade the reliability of the risk estimates.

Despite this recent evidence, previous work has been limited to simulations or single-method studies. It remains unclear how general these effects are across real-world clinical models and diverse algorithms. Numerous published risk models, across different healthcare domains, involve imbalanced outcomes and employ various machine learning techniques [22–29]. In this study, we aim to systematically re-evaluate the impact of applying commonly used class imbalance correction techniques on a range of published clinical prediction tasks. For each model and dataset, we compared discrimination and calibration as performance metrics with and without resampling on held-out test data or independent validation data.

## Materials and Methods

We conducted a cross-problem empirical evaluation to examine the impact of class imbalance correction on the performance of clinical risk prediction models addressing diverse problems in both in-hospital and out-of-hospital settings, including cardiometabolic diseases, sepsis, and mortality. Our approach involved implementing and retraining a heterogeneous set of models, including both previously described methods and newly developed models, using multiple published clinical datasets. We applied commonly used class-balancing techniques during model development and evaluated the resulting changes in discrimination and calibration on held-out test data, as illustrated in *Figure 1*.

To enhance reproducibility and representativeness, we invited the original authors of several published risk models to participate by retraining their respective models while modifying only the class distribution in the training data using standard resampling methods. All other aspects of the model development pipelines remained unchanged.

We hypothesized that (i) class imbalance correction methods would adversely affect model calibration, and (ii) such correction would not improve, and might even degrade model discrimination when evaluated on test data.

**Imbalance Correction Methods**

To examine the influence of class rebalancing on model performance, three commonly applied methods for handling class imbalance were implemented during model training: Random Undersampling (RUS), Random Oversampling (ROS), and the Synthetic Minority Oversampling Technique (SMOTE) [30,31]. RUS reduces the number of majority-class samples by randomly removing instances until a specified class ratio is achieved. ROS increases the representation of the minority class by randomly duplicating existing minority-class samples until class proportions are balanced. Lastly, SMOTE generates artificial minority-class samples by interpolating between existing minority instances in the feature space. This method aims to create a more continuous decision boundary between classes and reduce overfitting compared with random oversampling. For implementing SMOTE we used five neighbors (k=5). All resampling methods were applied exclusively to the training data, ensuring that the original class distribution in the test data remained unchanged for unbiased evaluation.

For each model, the degree of resampling was set to achieve a 1:1 class ratio in the training data. *Figure 2* illustrates the three class-imbalance correction methods using two features in a dataset with 200 cases and a class ratio of 8:2.

**Evaluation Metrics**

Model performance was assessed using standard discrimination and calibration metrics to quantify the effect of class imbalance correction.

Discrimination: The discrimination ability was assessed using ROC-AUC, which corresponds to the concordance statistics (c-statistics). The ROC-AUC estimates the probability that the model assigns a higher predicted probability to a randomly selected individual who experiences the event than to a randomly selected individual who does not. A ROC-AUC of 1 indicates perfect discrimination, where all cases with the event are assigned higher probabilities than all cases without the event, while an AUROC of 0.5 indicates no discriminative ability. ROC-AUC is valuable for comparing the diagnostic performance of multiple screening tests or evaluating the predictive ability of different models for the same condition. Generally, a model with a higher area under the curve (AUC) is considered to have superior discriminative performance [32,33]. In addition, we also report the Precision-Recall AUC (PR-AUC) as a complementary metric, which can provide additional insight, particularly in datasets with class imbalance [34].

Calibration metrics: Calibration refers to the agreement between predicted probabilities and observed outcomes, reflecting how well predicted risks correspond to actual event rates [35,36]. Calibration was evaluated using several complementary measures:

Calibration Slope: Indicates the degree of over- or underfitting in the model, with an ideal value of 1. A slope < 1 suggests overfitting (predictions too extreme), while a slope > 1 indicates underfitting (predictions too conservative).

Calibration Intercept: Reflects systematic over- or underestimation of risk, with an ideal value of 0. Positive values indicate consistent underestimation of risk, and negative values indicate consistent overestimation.

Brier Score: Measures the mean squared difference between predicted probabilities and observed outcomes, with lower values indicating better overall predictive accuracy [37]

Calibration Plots / reliability diagram: Graphically compare predicted and observed event rates across deciles of predicted risk, with perfect calibration represented by a diagonal (45°) line.

## Datasets

A diverse set of clinical datasets were included to evaluate the impact of class imbalance correction across different medical problems. All datasets contained binary outcomes and reflected both in-hospital and out-of-hospital scenarios. The selection aimed to represent clinically relevant risk prediction tasks including cardiometabolic diseases incidence and prevalence prediction and mortality risk.

Each dataset was obtained from publicly available sources or through collaborations with the authors of previously published risk models. The datasets varied in sample size, feature dimensionality/type, and baseline event rates, providing a broad representation of real-world clinical data. Here we briefly describe each dataset, for in-detail description we refer to the original publication(s) for each dataset and problem.

### Undiagnosed Type 2 Diabetes

The undiagnosed Type 2 Diabetes dataset was based on data drawn from the National Health and Nutrition Examination Survey (NHANES) spanning years 2005–2018 excluding pregnant women and individuals with a prior diabetes diagnosis. It included 45,431 participants aged over 20 years with 4.0% having undiagnosed Type 2 Diabetes (HbA1c ≥ 48 mmol/mol (6.5%)) and 96.0% without Type 2 Diabetes. The original study by S. Cichosz trained models with forward feature selection using clinically obtainable predictors (age, sex, ethnicity, BMI/waist circumference, blood pressure, sleep duration, physical activity, smoking, alcohol use, education and income-to-poverty ratio). We implemented this approach for an XG boost model and AutoGluon model using the dataset. [38]

**Weekly Hypoglycemia in Type 1 Diabetes**

The weekly hypoglycemia dataset was based on a primary cohort of long-term continuous glucose monitoring (CGM) data from 205 individuals with Type 1 Diabetes. The dataset included a total of 27,829 weeks of CGM data with 9.6% of the cases having time-below-range above 4% in the upcoming week and 90.4% having less than 4% in time-below-range. The original study by S. Cichosz trained models using features capturing glycaemic variability and hypoglycemia history (e.g., low blood glucose index, GRADE score, waveform length, coefficient of variation, and mean glucose levels from the previous week). We implemented this approach for an XG boost model and AutoGluon model using the dataset. [39]

**Elevated Ketones in Type 1 Diabetes**

The Elevated Ketones dataset consisted of 259 participants with Type 1 Diabetes aged 6–79 years who were using a closed-loop insulin delivery system, contributing over 49,000 days of CGM recordings and associated insulin data. In total, the dataset consisted of 1768 ketone measurements with 21.6% having elevated ketone body levels (≥ 0.6 mmol/L) and 78.4% without elevated levels. The original study by S. Cichosz trained models using a comprehensive set of engineered features capturing CGM dynamics, insulin delivery patterns, and self-monitored blood glucose metrics extracted over 6- and 12-hour windows prior to ketone measurements. We implemented this approach for a logistic regression model and a TabPFN model using the dataset. [40]

**Diabetes in Pima Indian Women**

The diabetes in Pima Indian Women dataset was based on the Pima Indians Diabetes Database, which is a widely used benchmark dataset in diabetes research, originally compiled by the National Institute of Diabetes and Digestive and Kidney Diseases. It consisted of 768 female participants of Pima Indian heritage aged 21 years and older with 35% having diabetes and 65% without diabetes. The original study by I. Abousaber trained different models using eight routinely collected health indicators, including number of pregnancies, plasma glucose concentration, diastolic blood pressure, triceps skinfold thickness, 2-hour serum insulin, body mass index, diabetes pedigree function and age. We implemented this approach for a logistic regression model and a CatBoost model using the dataset. [41]

**Survival in Sepsis patients**

The sepsis dataset was obtained from the Norwegian Patient Registry and linked with demographic records. It consisted of 110,204 hospital admissions including 84,811 individuals diagnosed with sepsis and related systemic inflammatory conditions infections between 2011 and 2012. In total, the dataset included 92.6% who survived during the hospital stay and 7.4% who died. The original study by D. Chicco trained a Gradient Boosting model using clinical features collected at hospital entry, including patient age, sex and count of sepsis episodes. We implemented this approach for a Gradient Boosting model using the dataset. [12]

**Self-reported Diabetes in U.S. adults**

The self-reported diabetes dataset was derived from the Centers for Disease Control and Prevention's Behavioral Risk Factor Surveillance System, which is an annual, nationwide telephone survey capturing health behaviors, chronic conditions, and preventive care utilization among non-institutionalized U.S. adults. It comprised 253,680 survey respondents with 13.9% having self-reported diabetes and 86.1% without diabetes. The original approach published on Kaggle by A. Teboul trained a feedforward artificial neural network (ANN) using 35 features encompassing demographic variables, clinical measures, lifestyle factors, healthcare access indicators, and self-reported health status. We implemented this approach for an ANN model using the dataset. [42,43]

**Cardiovascular Death in Prediabetes**

The prediabetes dataset was based on participants from a randomized controlled clinical trial (SELECT) [44,45], including only participants in the placebo group having prediabetes (HbA1c: 39-47 mmol/mol (5.7-6.4%)). In total, the dataset consisted of 5,636 individuals including 96.8% without CV death and 3.2% with CV death. The original study by A. Andersen developed a logistic regression model using seven clinical features: sex, age, high sensitive C-Reactive Protein (hsCRP), history of chronic heart failure event, alanine aminotransferase (ALT), gamma glutamyl transferase (GGT) and level of leucocytes. We implemented this approach for a Logistic regression model using the dataset.

**Coronary Artery Disease**

The Coronary Artery Disease (CAD) dataset was based on the Danish study of Non-Invasive Testing in Coronary Artery Disease (Dan-NICAD) cohort, which consists of prospective, multicenter, cross-sectional studies including patients presenting with de novo chest pain and no prior history of CAD [46–48]. The CAD dataset consisted of 4,407 individuals, comprising 11.1% cases with obstructive CAD, defined by invasive coronary angiography, and 88.9% cases without obstructive CAD. The original study by Winther et al. [49] evaluated the guideline-recommended risk factor-weighted clinical likelihood (RF-CL) risk score of obstructive CAD, which is based on clinical information including age, sex, chest pain characteristics, and cardiovascular risk factors [50]. The RF-CL model was validated in Dan-NICAD and guideline-recommended by the European Society of Cardiology in 2024 [51,52]. Furthermore, a model combining the RF-CL clinical risk score with a polygenic score (PGS) for CAD using publicly available GPSMult weights [53] were tested. We implemented this approach with and without the PGS for a logistic regression model using the dataset.

**Hypoglycemia in electronic Intensive Care Unit**

The hypoglycemia in eICU dataset was based on the ICU blood-glucose forecasting cohort constructed in [54], which is based on the eICU Collaborative Research Database (eICU-CRD) [55], a multi-center U.S. critical-care dataset (2014–2015). The dataset covered 200,859 ICU stays across 208 hospitals, including a total of 3,661,651 blood-glucose cases with 1.85% being hypoglycemia (blood-glucose <70 mg/dL) cases.

The original study by H. Mehdizavareh trained a Random Forest model to predict the next blood-glucose measurement using 50 engineered features following the feature schema in [56] (restricted to variables available in eICU). We implemented this approach for a Random Forest model using the dataset.

**48-hour Mortality in Intensive Care Unit**

The 48-hour mortality dataset was based on the retrospective single-centre cohort comprising 9,786 patients admitted between 2018 and 2022. The dataset included a total of 61,203 in hospital stay days with 4.6% belonging to the deceased class. The original study by S. Britsch trained a Gradient Boosting model to predict the 48-hour inhospital mortality updated every 24-hour in a tertiary medical intensive care unit (ICU) using routinely collected electronic health record (EHR) data including vital signs, laboratory values, and treatment parameters. We implemented this approach for a Gradient Boosting model using the dataset. [57]

**Statistical Analysis**

To assess the overall effect of class imbalance correction in our study for each performance metric, we included a combined effect analysis. Resampling approaches were compared against that of the original data. To estimate overall method performance, we used nonparametric bootstrap resampling. For each metric, 5,000 bootstrap samples were drawn with replacement from the empirical distribution of paired differences. For each bootstrap sample, the median difference was computed. The effect size was then taken as the mean of these bootstrap medians, providing a robust point estimate of the paired difference. Uncertainty in the effect estimate was quantified using the 2.5th and 97.5th percentiles of the bootstrap medians to define a 95% confidence interval (CI). To test whether the observed differences deviated significantly from zero, a two-sided Wilcoxon signed-rank test was applied to the paired differences for each metric. This nonparametric paired test was chosen because it does not assume normality of the differences and is robust to asymmetric distributions.

# Results

A total of ten clinical datasets spanning prediction instances from 605,842 patients in diverse medical domains were included in the analysis, covering a range of prediction tasks with outcome prevalences ranging from 1.9% to 34.9%. Across these datasets, multiple machine-learning model families were evaluated, including linear models such as logistic regression as well as non-linear approaches, including gradient boosting, CatBoost, RandomForest, artificial neural networks, XGBoost, and the tabular foundation model TabPFN. This resulted in a large set of model-domain combinations evaluated under four training conditions: models trained on the original data and models trained using three commonly applied 1:1 class-imbalance correction strategies. Individual model results are summarized in Table 1.

**Table 1. Model performance across datasets and class-imbalance correction strategies.**
Summary of discrimination and calibration metrics for multiple machine-learning models trained on original data and under three commonly used resampling methods across diverse clinical prediction tasks. Metrics include ROC-AUC (discrimination), PR-AUC, Brier score, calibration intercept, and calibration slope. The best performance for each metric for each model is highlighted in bold. For calibration intercept and slope, the absolute value closest to 0 and 1, respectively, is highlighted. † combined RF-CL score with a polygenic score (PGS) for coronary artery disease risk prediction.

| Dataset | Model | Method | ROC AUC | PR AUC | Brier | Intercept | Slope |
|---|---|---|---|---|---|---|---|
| Undiagnosed Type 2 Diabetes n=45,431 *Balance* (non-cases/cases) 96.0/4.0% | | | | | | | |
| | XGBoost | Original | **0.83** | **0.13** | **0.027** | **-0.20** | **0.96** |
| | XGBoost | ROS | **0.83** | 0.12 | 0.162 | -3.27 | 0.87 |
| | XGBoost | RUS | **0.83** | 0.12 | 0.184 | -3.42 | 0.82 |
| | XGBoost | SMOTE | 0.82 | 0.09 | 0.043 | -1.81 | 0.82 |
| | AutoGluon | Original | **0.84** | **0.14** | **0.027** | **0.20** | 0.98 |
| | AutoGluon | ROS | 0.82 | 0.13 | 0.029 | -1.36 | 0.62 |
| | AutoGluon | RUS | **0.84** | **0.14** | 0.183 | -3.51 | **1.01** |
| | AutoGluon | SMOTE | 0.73 | 0.06 | 0.031 | -2.04 | 0.17 |
| Weekly Hypoglycemia in Type 1 Diabetes n=27,829 *Balance* 90.4/9.6% | | | | | | | |
| | XGBoost | Original | **0.83** | **0.67** | **0.138** | **-0.12** | **0.96** |
| | XGBoost | ROS | **0.83** | 0.66 | 0.170 | -1.00 | 0.94 |
| | XGBoost | RUS | **0.83** | **0.67** | 0.170 | -1.02 | 0.94 |
| | XGBoost | SMOTE | **0.83** | 0.66 | 0.164 | -0.88 | 0.90 |
| | AutoGluon | Original | **0.80** | **0.63** | **0.160** | **-0.36** | 0.58 |

| | Model | Method | | | | | |
|---|---|---|---|---|---|---|---|
| | AutoGluon | ROS | **0.80** | 0.62 | 0.171 | 0.46 | 0.57 |
| | AutoGluon | RUS | **0.80** | 0.62 | 0.181 | -0.89 | **0.66** |
| Elevated Ketones in Type 1 Diabetes n=1768 Balance 78.4/21.6% | AutoGluon | SMOTE | 0.79 | 0.60 | 0.171 | -0.38 | 0.44 |
| | LogisticReg | Original | **0.76** | **0.46** | **0.144** | **-0.09** | **0.99** |
| | LogisticReg | ROS | 0.75 | 0.44 | 0.209 | -1.32 | 0.89 |
| | LogisticReg | RUS | 0.74 | 0.43 | 0.216 | -1.38 | 0.74 |
| | LogisticReg | SMOTE | 0.73 | 0.42 | 0.209 | -1.27 | 0.74 |
| | TabPFN | Original | **0.78** | **0.51** | **0.140** | 0.07 | **0.96** |
| | TabPFN | ROS | 0.75 | 0.44 | 0.171 | 0.61 | 0.69 |
| | TabPFN | RUS | **0.78** | 0.50 | 0.193 | -1.31 | 0.91 |
| Diabetes in Pima Indian Women n=768 Balance 65.1/34.9% | TabPFN | SMOTE | 0.77 | 0.49 | 0.158 | **0.02** | 0.48 |
| | LogisticReg | Original | **0.84** | **0.74** | **0.158** | **0.03** | 0.98 |
| | LogisticReg | ROS | **0.84** | **0.74** | 0.166 | -0.61 | **0.99** |
| | LogisticReg | RUS | **0.84** | **0.74** | 0.168 | -0.63 | 0.91 |
| | LogisticReg | SMOTE | **0.84** | **0.74** | 0.166 | -0.67 | 0.86 |
| | CatBoost | Original | **0.84** | **0.74** | 0.156 | 0.07 | 1.03 |
| | CatBoost | ROS | **0.84** | 0.72 | 0.164 | -0.44 | **1.01** |
| | CatBoost | RUS | 0.83 | 0.73 | 0.174 | 0.62 | 0.97 |
| Survival in Sepsis patients n= 110,204 Balance 92.6/7.4% | CatBoost | SMOTE | 0.82 | 0.71 | 0.170 | -0.42 | 0.86 |
| | GradientBoost | Original | **0.70** | **0.97** | **0.067** | **-0.46** | 1.19 |
| | GradientBoost | ROS | **0.70** | **0.97** | 0.225 | 2.53 | 1.18 |
| | GradientBoost | RUS | **0.70** | **0.97** | 0.226 | 2.54 | **1.12** |
| Self-reported Diabetes in U.S. adults n=253,680 Balance 86.1/13.9% | GradientBoost | SMOTE | **0.70** | **0.97** | 0.225 | 2.54 | 1.16 |
| | ANN | Original | **0.83** | **0.42** | **0.098** | **-0.09** | 0.91 |
| | ANN | ROS | 0.82 | 0.41 | 0.173 | -1.71 | 0.90 |
| | ANN | RUS | 0.82 | 0.41 | 0.183 | -1.83 | **0.98** |
| Cardiovascular Death in Prediabetes N=5636 Balance 97.3/2.7% | ANN | SMOTE | 0.80 | 0.37 | 0.160 | -1.47 | 0.62 |
| | LogisticReg | Original | **0.73** | **0.11** | 0.201 | -3.30 | **0.80** |
| | LogisticReg | ROS | **0.73** | **0.11** | 0.197 | -3.29 | 0.76 |
| | LogisticReg | RUS | 0.65 | 0.07 | 0.281 | -3.66 | 2.01 |
| Coronary Artery Disease n=4,407 Balance 88.9/11.1 % | LogisticReg | SMOTE | 0.62 | 0.07 | **0.176** | **-3.18** | 0.47 |
| | LogisticReg | Original | **0.76** | **0.29** | **0.089** | **0.05** | **1.02** |
| | LogisticReg | ROS | **0.76** | 0.28 | 0.201 | -2.08 | **0.98** |
| | LogisticReg | RUS | 0.75 | 0.28 | 0.212 | -2.12 | 0.90 |
| | LogisticReg | SMOTE | **0.76** | 0.28 | 0.198 | -2.05 | 0.90 |
| | LogisticReg† | Original | **0.81** | **0.35** | **0.085** | **0.15** | 1.09 |
| | LogisticReg† | ROS | **0.81** | **0.35** | 0.189 | -2.12 | 1.06 |
| | LogisticReg† | RUS | 0.79 | 0.34 | 0.200 | -2.15 | 0.93 |
| Hypoglycemia in electronic Intensive Care Unit n=3,661,651 Balance 98.2/1.8 % | LogisticReg† | SMOTE | 0.80 | 0.34 | 0.186 | -2.09 | **0.95** |
| | RandomForest | Original | 0.88 | **0.22** | **0.016** | -2.73 | 0.11 |
| | RandomForest | ROS | 0.86 | 0.20 | 0.017 | **-2.68** | 0.09 |
| | RandomForest | RUS | **0.91** | 0.19 | 0.118 | -3.66 | **0.56** |
| | RandomForest | SMOTE | 0.88 | 0.19 | 0.020 | -2.72 | 0.27 |
| 48-hour Mortality in Intensive Care Unit n=61,203 Balance 95.4/4.6 % | GradientBoost | Original | **0.89** | **0.43** | **0.034** | **-0.05** | 0.87 |
| | GradientBoost | ROS | 0.88 | 0.42 | 0.037 | -0.91 | 0.66 |
| | GradientBoost | RUS | 0.88 | 0.39 | 0.155 | -3.16 | **0.91** |
| | GradientBoost | SMOTE | 0.88 | 0.42 | 0.35 | -0.36 | 0.71 |

**Discrimination**

Overall discrimination performance, quantified by ROC-AUC, was similar or worse across models trained using resampling techniques compared with models trained on the original data. Only in one case (Hypoglycemia in electronic Intensive Care Unit) did RUS improve ROC-AUC slightly from 0.88 to 0.91. This case is interesting because with the large sample size (n=3,661,651), RUS might reduce the dominance of majority patterns and alter tree splits in a way that improves ranking. However, it comes with a large Brier increase (0.016 to 0.118) and lower PR-AUC (0.22 to 0.19). Across most datasets and algorithms, ROC-AUC values differed only marginally between training using class imbalance strategies and no resampling approach. As an example, Figure 3 illustrates the differences between the different training strategies for predicting hypoglycemia.

When aggregating results across all datasets and model families, Figure 4, the median change in ROC-AUC relative to the original models was small for ROS and SMOTE (-0.002 to -0.01, p<0.05). For RUS, no difference was observed (p=0.12). In addition, changes in PR AUC relative to models trained on the original data were in general small but lower (ROS: -0.1, p<0.01; RUS: -0.1, p<0.01; SMOTE: -0.03, p<0.01), showing that resampling did not improve the precision–recall performance.

**Calibration**

In contrast to discrimination, calibration was more strongly affected by resampling. Across multiple datasets, models trained with resampling exhibited worse probabilistic accuracy, as reflected by higher Brier scores compared with models trained on the original data. Deviations in calibration intercept and calibration slope were frequently observed following resampling, indicating systematic shifts in predicted risk levels. This is further exemplified in Figure 3, illustrating the calibration across different training strategies for hypoglycemia prediction.

These calibration effects were evident across different resampling strategies and model families and were not confined to a single dataset or clinical domain. In several tasks, resampling resulted in substantial overestimation or underestimation of absolute risk, despite minimal changes in ROC-AUC. This pattern highlights a dissociation between discrimination and calibration, whereby rank-based performance was preserved while the reliability of predicted probabilities deteriorated.

When aggregating results across all datasets and model families, Figure 4, the median change in Brier score relative to the original models was higher for all resampling strategies (0.029 to 0.080, p<0.05). Effect on intercept and slope was generally variating between prediction task and training strategy, but a tendency to both lower slope and intercept was observed.

# Discussion

In this study, we systematically re-evaluated a range of published clinical prediction tasks to quantify how common class-imbalance corrections applied during training affect both discrimination and calibration on held-out data. Across heterogeneous datasets and model families, the principal finding was consistent: resampling produced at best negligible changes in discrimination (ROC-AUC) but frequently produced substantially worse probabilistic calibration. This indicated that resampling might harm the risk probability estimates, which are important for clinical decision-making, even without improving the model's discriminative ability. These results confirm prior simulation studies [14,21] and extend them to real clinical datasets and multiple algorithms.

In our study, resampling did not improve ROC-AUC for most tasks, and several resulted in reduced predictive ability. By contrast, the Brier score often increased (worsened) after resampling. Calibration intercepts and slopes frequently moved away from ideal values (intercept = 0, slope = 1), sometimes dramatically. These patterns show that resampling can distort the mapping between model outputs and true outcome probabilities even when discrimination is unchanged.

Together, the results support these practical points. First, if the primary goal is to obtain well-calibrated risk probabilities for decision support, routine rebalancing of the training set is not recommended because it commonly degrades calibration. Since resampling changes how frequent each class is seen during training, the model learns probabilities from an altered event rate. When deployed on the true event rate in the held-out test data, the probabilities could systematically be shifted up or down, resulting in a miscalibrated model unless the altered event rate is corrected. Undersampling could mitigate model bias toward the majority class but may also lead to information loss, particularly when the number of minority events is very small. Furthermore, while oversampling retains all majority-class information, it may increase the risk of overfitting by repeatedly exposing the model to identical minority instances. Compared with random oversampling, SMOTE might be better in reducing overfitting. However, SMOTE may introduce unrealistic synthetic samples, particularly in high-dimensional or heterogeneous feature spaces. E.g. SMOTE can create non-physical or non-realistic combinations (e.g., interpolating categorical or constrained laboratory values) in clinical tabular data. This could impact the learned patterns and contribute to calibration harm, that are difficult to correct by recalibration. Second, many benefits attributed to resampling, such as improved sensitivity or accuracy at a chosen cutoff, can alternatively be achieved by post-hoc decision threshold tuning on a calibrated model without changing the underlying probability estimates. Whenever classifiers are used to provide probabilities, calibration must be evaluated and reported, and recalibration on a validation set should be considered before deployment.

**Strengths and limitations**

This study has several important strengths. First, it extends prior simulation-based work by evaluating the impact of resampling techniques across a diverse set of real-world clinical prediction tasks and multiple machine learning algorithms, encompassing ten datasets with more than 600,000 patients and eight different model types, including both linear and non-linear methods. This approach increases the validity of the findings and demonstrates that the observed calibration effects are not confined to synthetic or narrowly defined settings. Second, by focusing explicitly on calibration we capture performance changes that are often overlooked when only ROC-AUC or threshold-based performance metrics are reported. Finally, the use of held-out test or independent validation data ensures that the reported calibration differences reflect genuine out-of-sample behavior.

Some limitations should also be acknowledged. The study evaluated three commonly used resampling strategies. We only tested 1:1 resampling, but softer target ratios (e.g. 1:4) might reduce the distortion of the probabilistic calibration while still impacting the learning behavior. Further, other approaches including hybrid resampling were not examined and may exhibit different properties. The datasets themselves vary considerably in size, outcome prevalence, and data size both regarding feature complexity and instances - while this heterogeneity strengthens generalizability, it also means that the behavior of resampling in our study may not generalize to extremely small, highly imbalanced datasets or datasets with fundamental difference properties of predictors. Finally, although we documented substantial degradation in calibration following resampling, we did not comprehensively evaluate all available recalibration techniques. Future work could explore the effect of recalibration and whether recalibration performance differs across different resampling methods and model complexity. Therefore, it remains possible that certain combinations of methods and dataset could benefit from resampling.

**Conclusion**

In a collection of diverse real clinical prediction tasks, commonly used training-time imbalance correction techniques (random oversampling, random undersampling and SMOTE) did not improve discrimination and frequently degraded probabilistic calibration. Given that accurate probability estimates are essential for many clinical applications, it is advised not to apply resampling by default when calibrated risks are important. Instead, it is recommend: (1) to train on the original data whenever possible, (2) to evaluate and report calibration, (3) to use threshold tuning or cost-sensitive decision rules to achieve the desired sensitivity/specificity operating points, and (4) if resampling is used, validate calibration on independent data and apply recalibration before clinical use.

**Funding**: This work was supported by a research grant to A.K.A from the Danish Diabetes and Endocrine Academy, funded by the Novo Nordisk Foundation grant number NNF22SA0079901, and the Danish Data Science Academy, funded by the Novo Nordisk Foundation grant number NNF21SA0069429.

**Conflicts of Interest**: A.K.A. is a Novo Nordisk employee and has received a PhD grant from the Danish Diabetes and Endocrine Academy (grant number NNF22SA0079901) and the Danish Data Science Academy (grant number NNF21SA0069429), funded by the Novo Nordisk Foundation. H.M. and A.K. acknowledge support from the Novo Nordisk Foundation grant NNF 22OC0072415. T.B. is an employee and shareholder of Roche Diagnostics International, Totkreuz, Switzerland. M.B. disclose advisory board participation for Astra-Zeneca, Bayer, Boehringer Ingelheim, Novartis, NOVO Nordisk and Sanofi. S.W. were supported by a grant from the Novo Nordisk Foundation (grant number NNF21OC0066981). M.H.J. is employed at Novo Nordisk A/S and holds shares in Novo Nordisk A/S. S.L.C has received research funding from i-SENS Inc., holds shares in Novo Nordisk, and has received consultancy fees from Roche Diagnostics. The remaining authors declare that there are no conflicts of interest in this work.

**Data availability statement**: The majority of raw data utilized in this study are not publicly available due to the inclusion of sensitive patient information, which is subject to strict confidentiality and privacy regulations. Access to the data is restricted to ensure compliance with ethical guidelines and to protect patient privacy. Requests for additional information or collaboration may be considered on a case-by-case basis, subject to appropriate ethical approval and data-sharing agreements. For additional information on the separate dataset/studies we refer to the original publications referenced in the method section.

**Ethics Statement**: The presented study is a reanalysis of existing and anonymized data from several clinical trials and public available datasets. The original study protocols and informed consent forms were approved by the institutional review board prior to each enrollment of patients. For additional information on the separate dataset/studies we refer to the original publications referenced in the method section.

**Author contributions**

Amalie Koch Andersen (Conceptualization, Data Curation, Formal analysis, Methodology, Visualization, Writing - original draft, Writing - review & editing), Hadi Mehdizavareh (Data Curation, Formal analysis, Methodology, Writing - review & editing), Arijit Khan (Writing - review & editing), Tobias Becher (Data Curation, Formal analysis, Methodology, Writing - review & editing), Simone Britsch (Data Curation, Formal analysis, Methodology, Writing - review & editing), Markward Britsch (Data Curation, Formal analysis, Methodology, Writing - review & editing), Morten Bøttcher (Formal analysis, Methodology, Writing - review & editing), Simon Winther (Data Curation, Formal analysis, Methodology, Writing - review & editing), Palle Duun Rohde (Data Curation, Formal analysis, Methodology, Writing - review & editing), Morten Hasselstrøm

**Figure legends**

**Figure 1. Method Overview**
Each of the ten imbalanced clinical datasets (left) was used to train machine learning models under four conditions: original data and three commonly applied class-imbalance correction strategies: Random Undersampling (RUS), Random Oversampling (ROS), and Synthetic Minority Oversampling Technique (SMOTE). Multiple model families, including linear and non-linear approaches, were trained and validated using cross-validation. Performance was assessed on held-out data (internal validation) using discrimination metrics (ROC-AUC) and calibration metrics (Brier score, calibration slope and intercept). The analysis quantified the effect of imbalance correction on both discrimination and probabilistic calibration across diverse clinical prediction tasks.

**Figure 2. Visual illustration of class-imbalance correction strategies.**
Two-dimensional feature space showing the distribution of majority (blue) and minority (orange) classes under four conditions: Original (top-left), Random Undersampling (RUS) (top-right), Random Oversampling (ROS) (bottom-left), and Synthetic Minority Oversampling Technique (SMOTE) (bottom-right). RUS reduces the majority class to match the minority class size, ROS duplicates minority samples which are illustrated by a larger point size, and SMOTE generates synthetic minority samples by interpolating between existing minority points.

**Figure 3. Calibration performance of XGBoost for weekly hypoglycemia risk prediction**
Calibration plots compare predicted probabilities (x-axis) against observed frequencies (y-axis) for models trained on original data (top-left), Random Oversampling (ROS) (top-right), Random Undersampling (RUS) (bottom-left), and Synthetic Minority Oversampling Technique (SMOTE) (bottom-right). The dashed red line represents perfect calibration. Each panel reports Brier score, calibration intercept, slope, and ROC-AUC. While discrimination (ROC-AUC = 0.83) remained stable across all strategies, resampling methods substantially distorted calibration, with higher Brier scores and slopes deviating from 1, indicating systematic miscalibration despite preserved rank-based performance.

**Figure 4. Combined analysis of the effect of class-imbalance correction methods on model performance and calibration metrics.**
Forest plots show pooled differences (with 95% confidence intervals represented by the error bars) between models trained with resampling strategies - Random Undersampling (RUS), Random Oversampling (ROS), and Synthetic Minority Oversampling Technique (SMOTE) - relative to models trained on original data across multiple clinical prediction tasks. Metrics include ROC-AUC (top-left), Brier score (top-right), Calibration intercept (bottom-left), and Calibration slope (bottom-right). For ROC AUC and Brier score, the preferred direction is indicated by an arrow, corresponding to higher ROC AUC values and lower Brier scores, respectively. For Intercept and Slope, the ideal values are defined relative to the median differences between the model trained on the original data and the theoretical optimal values of 0 and 1, respectively.

**Figure 1** Method Overview

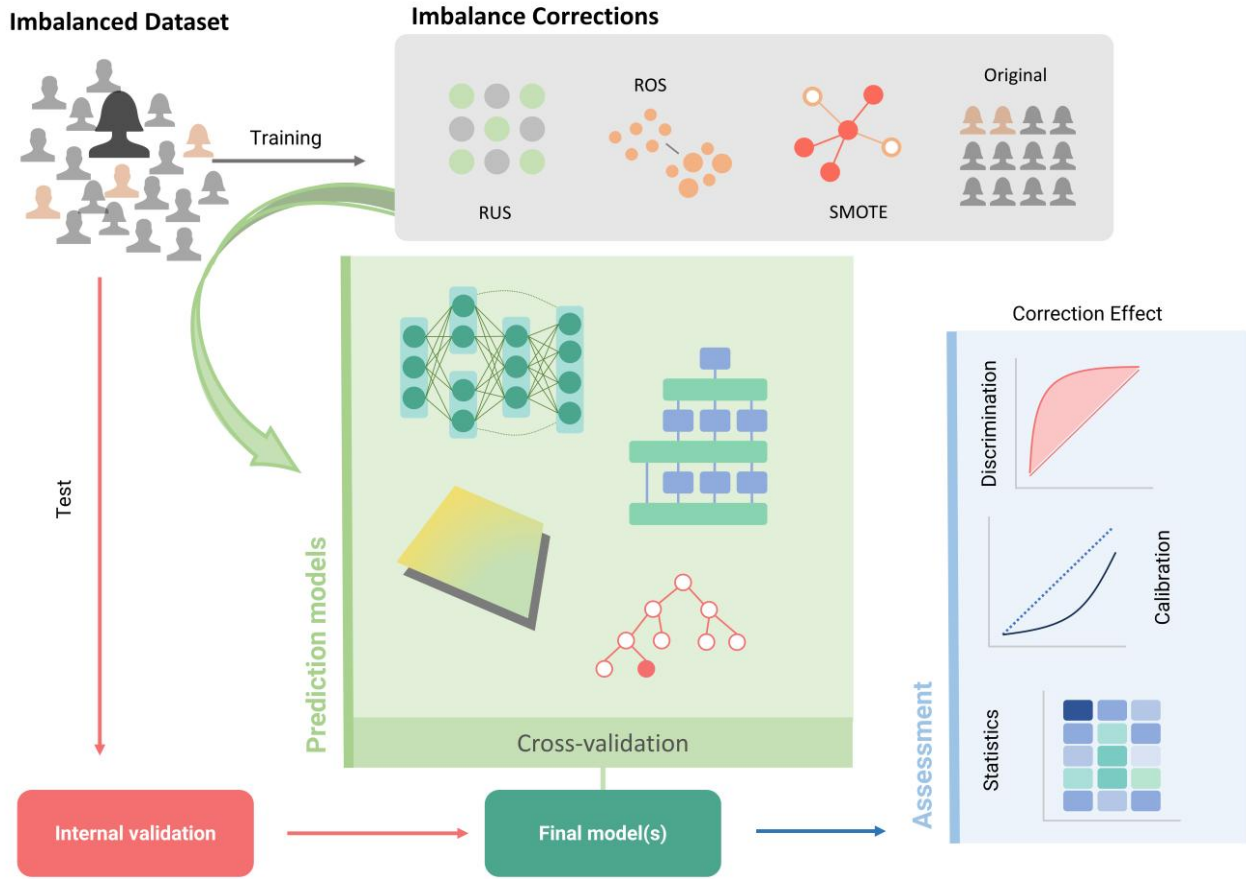

**Figure 1. Method Overview**
Each of the ten imbalanced clinical datasets (left) was used to train machine learning models under four conditions: original data and three commonly applied class-imbalance correction strategies: Random Undersampling (RUS), Random Oversampling (ROS), and Synthetic Minority Oversampling Technique (SMOTE). Multiple model families, including linear and non-linear approaches, were trained and validated using cross-validation. Performance was assessed on held-out data (internal validation) using discrimination metrics (ROC-AUC) and calibration metrics (Brier score, calibration slope and intercept). The analysis quantified the effect of imbalance correction on both discrimination and probabilistic calibration across diverse clinical prediction tasks.

**Figure 2** Class Imbalance Correction Methods

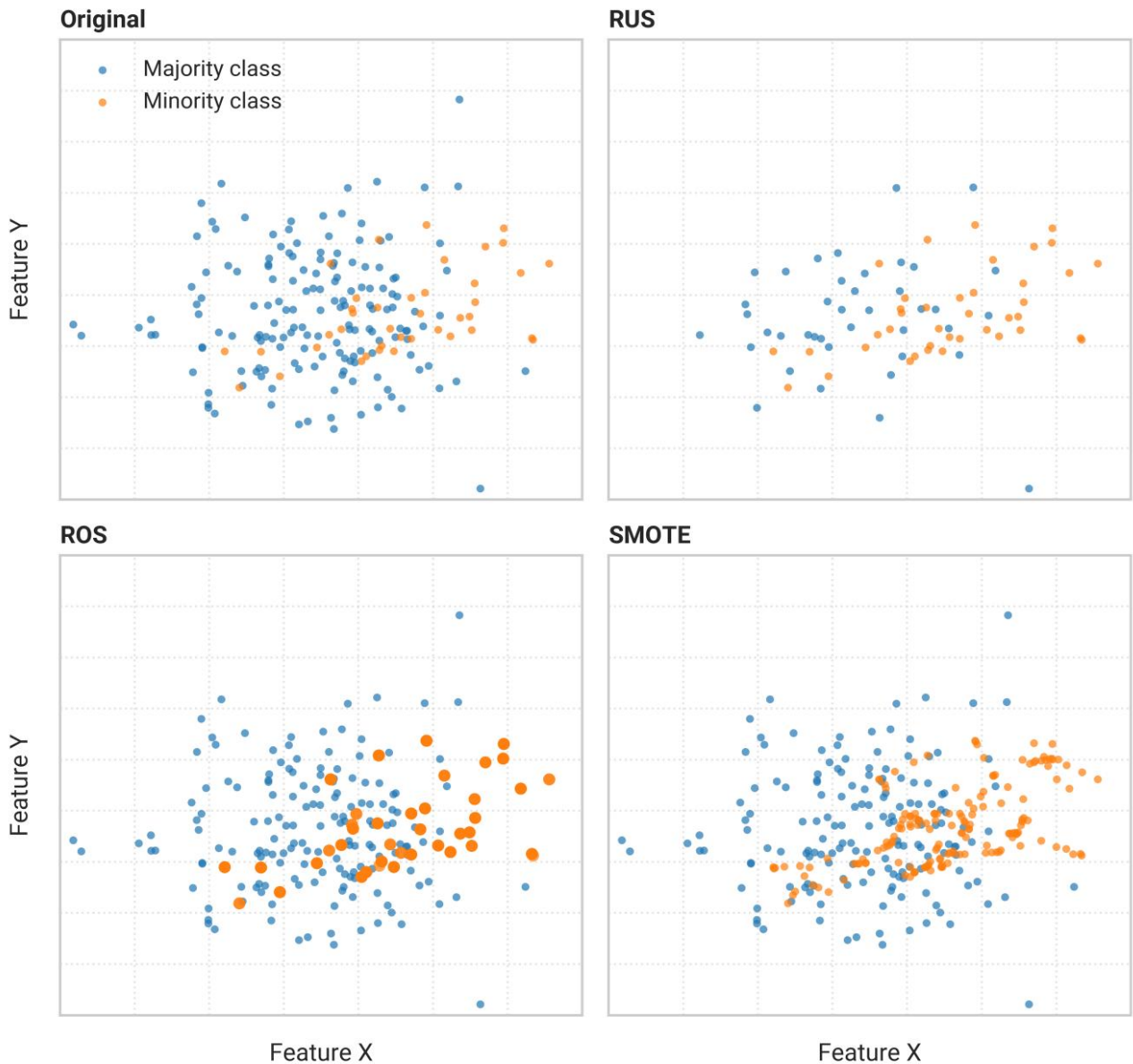

**Figure 2. Visual illustration of class-imbalance correction strategies.**
Two-dimensional feature space showing the distribution of majority (blue) and minority (orange) classes under four conditions: Original (top-left), Random Undersampling (RUS) (top-right), Random Oversampling (ROS) (bottom-left), and Synthetic Minority Oversampling Technique (SMOTE) (bottom-right). RUS reduces the majority class to match the minority class size, ROS duplicates minority samples which are illustrated by a larger point size, and SMOTE generates synthetic minority samples by interpolating between existing minority points.

**Figure 3** Calibration XGBoost: Weekly Hypoglycemia Risk

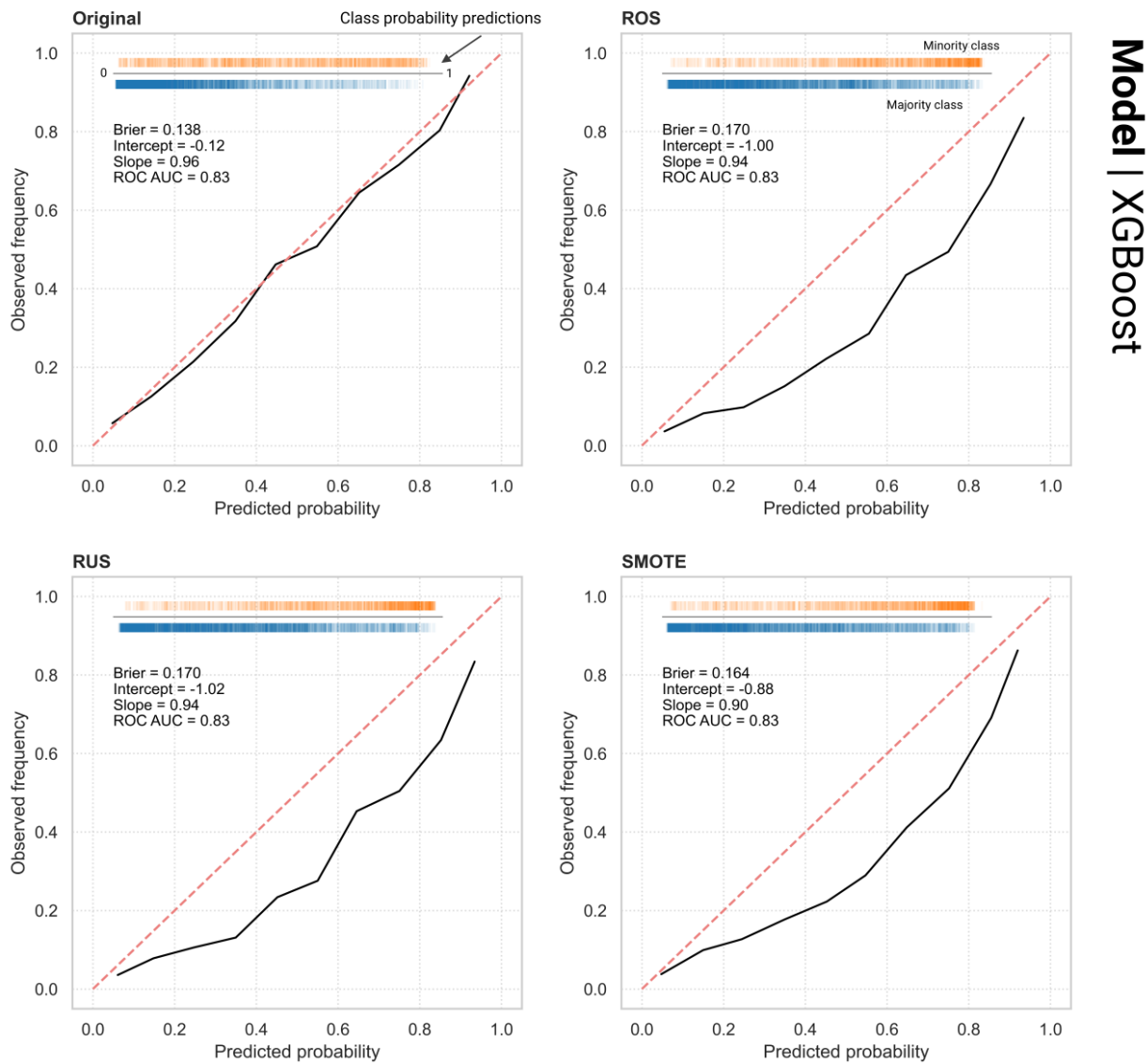

**Figure 3. Calibration performance of XGBoost for weekly hypoglycemia risk prediction**
Calibration plots compare predicted probabilities (x-axis) against observed frequencies (y-axis) for models trained on Original data (top-left), Random Oversampling (ROS) (top-right), Random Undersampling (RUS) (bottom-left), and Synthetic Minority Oversampling Technique (SMOTE) (bottom-right). The dashed red line represents perfect calibration. Each panel reports Brier score, calibration intercept, slope, and ROC-AUC. While discrimination (ROC-AUC = 0.83) remained stable across all strategies, resampling methods substantially distorted calibration, with higher Brier scores and slopes deviating from 1, indicating systematic miscalibration despite preserved rank-based performance.

**Figure 4** Combined Analysis: Effect of Methods

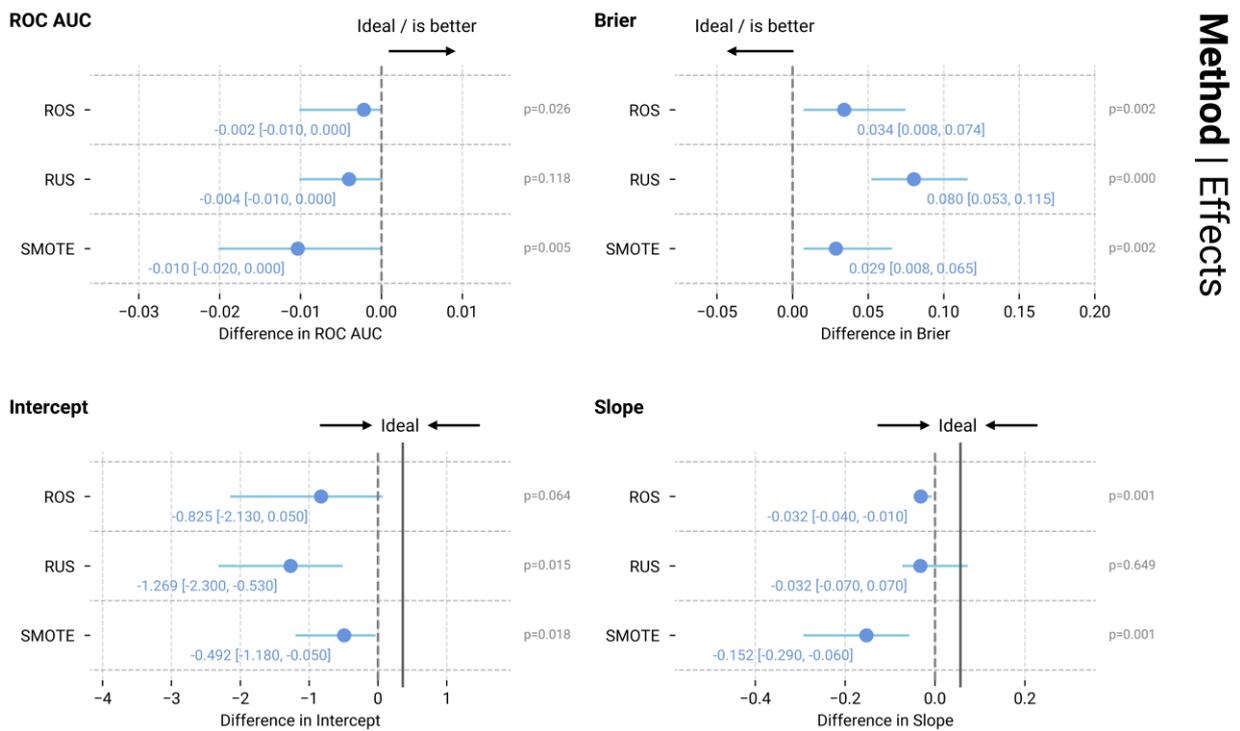

**Figure 4. Combined analysis of the effect of class-imbalance correction methods on model performance and calibration metrics.**

Forest plots show pooled differences (with 95% confidence intervals represented by the error bars) between models trained with resampling strategies - Random Undersampling (RUS), Random Oversampling (ROS), and Synthetic Minority Oversampling Technique (SMOTE) - relative to models trained on original data across multiple clinical prediction tasks. Metrics include ROC-AUC (top-left), Brier score (top-right), Calibration intercept (bottom-left), and Calibration slope (bottom-right). For ROC AUC and Brier score, the preferred direction is indicated by an arrow, corresponding to higher ROC AUC values and lower Brier scores, respectively. For Intercept and Slope, the ideal values are defined relative to the median differences between the model trained on the original data and the theoretical optimal values of 0 and 1, respectively.